\documentclass[aps,pra,amsmath,amssymb,showpacs,reprint]{revtex4-1}

\usepackage{subfigure}
\usepackage[breaklinks=true]{hyperref}
\usepackage{graphicx}
\usepackage{color}
\newcommand{\bra}[1]{\langle #1|}
\newcommand{\ket}[1]{|#1\rangle}

\newcommand{\ii}{\textrm{i}}

\newcommand{\tr}{\textrm{tr}}

\begin{document}

\title{Steady-state entanglement enhanced by a dissipative ancilla}
\author{Joachim Fischbach}
\email{joachim.fischbach@uni-ulm.de}
\author{Matthias Freyberger}
\affiliation{Institut f\"ur Quantenphysik and Center for Integrated Quantum
Science and Technology ($IQ^{ST}$), Universit\"at Ulm, D-89069 Ulm, Germany}
\date{\today}

\begin{abstract}

We investigate how to enhance entanglement in the steady state of interacting
two-level systems. The steady state is reached by spontaneous decay of the
individual systems. When we additionally couple them to a dissipative two-level
ancilla with variable eigenfrequency and coupling strength, we observe a
considerable enhancement effect in the entanglement of this steady state.
Moreover, we see that the increased entanglement is directly connected to the 
selection of certain excited states via the environment disturbing the ancilla.
This effect could be used in dissipative state preparation schemes as well as a
testbed for decoherence models.

\end{abstract}

\pacs{03.67.Bg, 03.65.Yz}

\maketitle

\section{Introduction}

Entanglement is a very peculiar feature and one of the hall marks of quantum
mechanics \cite{Einstein1935, Schroedinger1935}.
Its fundamental awkwardness, that surfaces for example in tests of the Bell
inequalities \cite{Bell1964, Aspect1982}, has been subject to scientific
discussion since the early days of quantum mechanics. Besides these conceptual
difficulties in grasping the notion of entanglement, in present-day quantum
information theory \cite{Nielsen2000}, it is considered as a resource
\cite{Horodecki2009}. Many stunning achievements, like quantum teleportation
\cite{Bennett1993}, quantum cryptography \cite{Ekert1991}, quantum simulators
\cite{Lloyd1996} or quantum computational algorithms
\cite{Deutsch1992,Shor1994,Grover1996} rely on entangled quantum states more or
less heavily.

Usually, the awkwardness of the quantum world stays hidden from our every day
observations. The belief is, that environmental decoherence destroys
counterintuitive quantum properties, like entanglement. This transition from the
quantum to the classical world is one of the central achievements in the theory
of decoherence \cite{Zurek1981,Joos1985,Zurek2003}. According to this theory,
the quantum state of every system in contact with an environment ultimately
decays into a classical mixture of states, where the explicit form of these
states is governed by the system reservoir interaction. 

From this perspective it seems rather conflicting, that environmental
decoherence also has been found to be useful \cite{Plenio1999}. In special
systems entanglement can be created by decoherence
\cite{Plenio2002,Schneider2002,Braun2002} or even quantum computations may be
performed dissipatively \cite{Verstraete2009}. The basic idea of engineering
dissipative environments to prepare interesting quantum states 
\cite{Poytas1996} has been extended to the preparation of entangled quantum
states \cite{Kraus2008,Verstraete2009,Schirmer2010}. 
Nowadays, a plethora of dissipative quantum state preparation schemes has
emerged, where the steady state of the system shows the peculiar feature of
quantum entanglement, despite or rather just because being subject to the
influence of environmental decoherence. In most of them the environment is
engineered by applying active driving of the systems by microwaves or lasers.
Such active schemes exist for many different physical systems, like ions in
traps \cite{Barreiro2011,Lin2013,Cormick2013}, atoms in cavities
\cite{Paternostro2004,Kraus2004,SZhang2009,Kastoryano2011,Reiter2012,Chen2012,
Sweke2013}, ensembles of atoms
\cite{Krauter2011,Muschik2011,Parkins2006,Dalla_Torre2013}, directly coupled
solid state qubits \cite{Zhang2009,Xia2011} and solid state qubits in resonators
\cite{Shen2011,Reiter2013,Leghtas2013,Shankar2013,Aron2014,Hou2014}. However,
there are also passive schemes
\cite{Arnesen2001,Hartmann2006,Zhang2009,Xia2011,Huelga2012,Sauer2014}, where
the steady state of the system can be reached 
without external driving. In these schemes the interaction between the systems
constituents ensures an entangled steady state.

\begin{figure}[htb!]
 \includegraphics{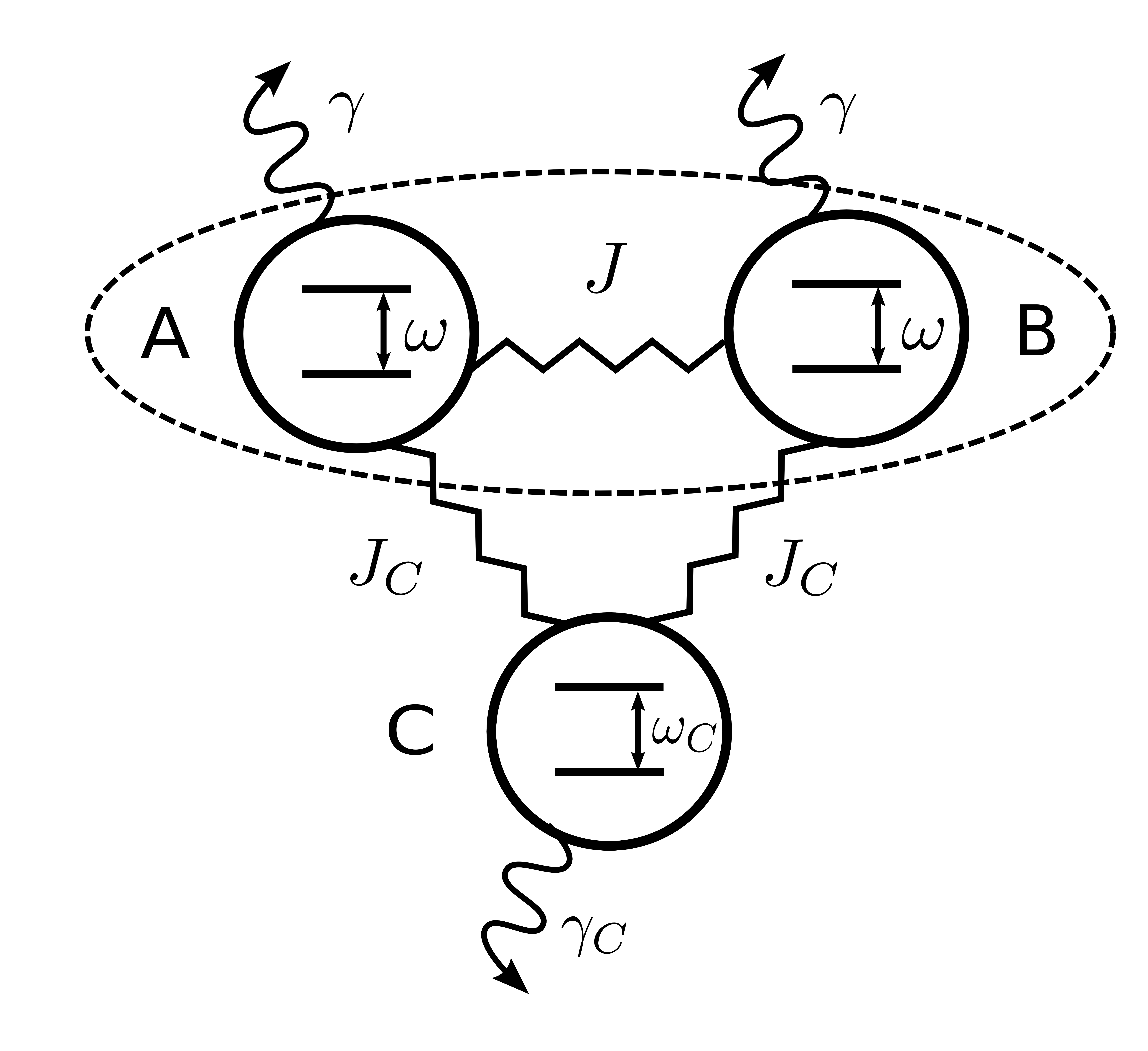} 
 \caption{We are interested in the steady state entanglement between systems $A$
and $B$ and especially how this entanglement depends on the properties of the
dissipative ancilla $C$. The model consists of three two-level systems $A, B$
and $C$ with eigenfrequencies $\omega$ and $\omega_C$ that are coupled
symmetrically, where $J$ and $J_C$ are the coupling strengths. Each of the
systems is located in a heat bath, characterized by the decoherence parameters
$\gamma$ and $\gamma_C$, respectively.}
 \label{sysschematic}
\end{figure}

In this manuscript we also examine a passive scheme. We are interested in how
steady state entanglement can be enhanced in the simplest bipartite system.
Therefore we extend the models of two coupled two-level systems in heat baths,
studied for example in Refs. \cite{Zhang2009,Sauer2014}, by a third two-level
system $C$, that mediates the coupling to an additional heat bath, see Fig.
\ref{sysschematic}.
One might say, that this dissipative ancilla, this means system $C$ and its heat
bath, forms an interesting, engineered environment with regard to systems $A$
and $B$. We refer to it as engineered, because we assume to have some control
over system $C$, giving us the chance to manipulate the dissipative ancilla. In
this sense controllable two-level systems appear for example in experiments with
artificial atoms in superconducting solid state systems
\cite{Bouchiat1998,Nakamura1999}, where parameters like the eigenfrequency can
be tuned. Due to this engineering of an environment we rely on the foundational
principles of dissipative state preparation
\cite{Poytas1996,Kraus2008,Verstraete2009,Schirmer2010}. However, in our work we
start from an experimentally motivated and very limited setup. In particular, we
assume that all coupling operators in the Hamiltonian and the Lindblad operators
describing our model of decoherence are fixed. We only vary coupling parameters
and eigenfrequencies. In this way, we obtain a practical scheme for preparing
entangled states or storing entanglement in presence of a decohering 
environment.

An additional peculiarity is the treatment of the coupling between our
dissipative ancilla $C$ and the systems $A$ and $B$ without any further
approximations. Hence we preserve the full quantum dynamics of this interaction,
relating our approach to studies of environmentally enhanced entanglement as
performed in Ref. \cite{Huelga2012} for a different decoherence mechanism. Last
but not least, the predicted enhancement effect clearly relies on the
decoherence model we utilize. This opens up the possibility to sensitively test
the phenomenological modeling of decoherence by experimentally checking for
such an intricate quantum effect.

Our paper is organized as follows. We start by detailing our model of bipartite
system and dissipative ancilla in section \ref{Model}. In section \ref{Result},
we study how to find parameters of the ancilla to enhance the steady state
entanglement via dissipation. In section \ref{dsoe} we connect this effect to
the dissipative preparation of energy eigenstates. This further allows us
to understand the optimum of the enhancement. Finally, we arrange our findings
and conclude with section \ref{Conclusion}.

\section{Model}
\label{Model}

Our model consists of three two-level systems $A, B$ and $C$, that are coupled
symmetrically by a $\sigma_x \sigma_x$ interaction, see Fig. \ref{sysschematic}.
This form of interaction is often found in systems involving artificial atoms,
see for example Ref. \cite{Zhang2009}. All of them are located in thermal baths.
In this way we model spontaneous decay. We choose $A$ and $B$ to be equal, i.e.
having the same eigenfrequencies $\omega_A = \omega_B = \omega $. Also, these
two systems are coupled to the third system with the same interaction strength
$J_C$. This construction allows us to study the entanglement between the two
systems $A$ and $B$ and how it depends on the properties of the dissipative
ancilla $C$. 
The Hamiltonian $H = H_f + H_{int}$ of all three systems combines the free
Hamiltonian
\begin{equation}
 H_f = \frac{\hbar \omega}{2} \left( \sigma_z^A + \sigma_z^B \right) +
\frac{\hbar \omega_C}{2} \sigma_z^C , \label{Hfree}
\end{equation}
that describes the free evolutions of the systems $A,B$ and $C$ with their
corresponding eigenfrequencies $\omega$ and $\omega_C$
and the interaction part
\begin{equation}
 H_{int} = \hbar J \sigma_x^A \sigma_x^B + \hbar J_C \left( \sigma_x^A
\sigma_x^C + \sigma_x^B \sigma_x^C  \right) , \label{Hinteraction}
\end{equation}
which realizes the coupling with strength $J$ between $A$ and $B$ and coupling
strength $J_C$ between $A$ and $C$ as well as $B$ and $C$. In this simplest
version of our model, we assume zero temperature heat baths, that are modeled by
a coupling of the individual systems to a continuum of harmonic oscillators
\cite{Breuer2002}. After applying the standard Born-Markov approximations, one
ends up with a dissipator in Lindblad form \cite{Kossakowski1972,Lindblad1976}
\begin{equation}
 \mathcal{L}(\rho) = \sum\limits_{k \in \{A,B,C\}} \gamma_k \left( \sigma_{-}^k
\rho \sigma_{+}^k - \frac{1}{2} \sigma_{+}^k \sigma_{-}^k \rho - \frac{1}{2}
\rho \sigma_{+}^k \sigma_{-}^k  \right) ,
\end{equation}
where $\sigma_{-}^k = \ket{g}\bra{e}_k$ is the Lindblad operator that describes
spontaneous decay from the excited level $\ket{e}_k$ into the ground state
$\ket{g}_k$ of system $k$ and  $\sigma_{+}^k$ denotes its adjoint. For
simplicity, we assume symmetric decoherence parameters $\gamma_A = \gamma_B =
\gamma$.
The full master equation of our model thus reads
\begin{equation}
 \dot{\rho} = L\left[ \rho \right] = - \frac{\ii}{\hbar} \left[ H , \rho \right]
+ \mathcal{L}(\rho) . \label{louivillian}
\end{equation}
Before solving it, we introduce the dimensionless time $\tilde{t}=\omega t$,
which effectively rescales all eigenfrequencies, coupling strengths and
decoherence parameters $x$ by the eigenfrequency $\omega$ of the two fundamental
systems, i.e. $\tilde{x} = x / \omega$. We drop all tildes and continue with
these dimensionless variables.

The steady state solution $\dot{\rho}_{st}=0$ of Eq. (\ref{louivillian}) is
unique due to the presence of the decay operators $\sigma_{-}$ in every
subsystem \cite{Kraus2008,Schirmer2010}. We find it by solving $L\left[
\rho_{st} \right] = 0$ numerically \footnote{Solving the algebraic equation
$L[\rho] = 0$ is equivalent to determining the eigenvector of the linear map
$L[\cdot]$ associated with eigenvalue zero. A matrix representation of terms $A
\rho B$ in the Liouvillian $L[\cdot]$ is given by $A \otimes B^{\dagger} .
\vec{\rho}$, where $\vec{\rho}$ is the column vector constructed by appending
the rows of the density matrix $\rho$ and transposing the result.
We make use of the qutip library \cite{Johansson2012,Johansson2013}, that
provides several out of the box tools to solve master equations in Lindblad
form.
}. To measure the entanglement between systems $A$ and $B$, we calculate the
Negativity \cite{Zyczkowski1998,Vidal2002}
\begin{equation}
 N(\rho_{AB}) = \frac{1}{2} \left(
\left|\left|\left(\rho_{AB}\right)^{T_B}\right|\right| - 1  \right)
\label{negdef}
\end{equation}
of the reduced steady state density matrix $\rho_{AB} = \tr_C [ \rho_{st} ]$,
where $T_B$ means partially transposed with respect to subsystem $B$ and $|| X
|| = \tr[\sqrt{X^{\dagger}X}]$ is the trace norm.

The model we have chosen is simple, consisting only of three coupled two-level
systems in Markovian heat baths. This will allow us to study the influence of
tunable parameters like eigenfrequency $\omega_C$ or coupling strength $J_C$ of
the dissipative ancilla $C$ on the steady state entanglement $N(\rho_{AB})$
between $A$ and $B$. In addition, such variable eigenfrequencies and coupling
strengths immediately lead us to think of experiments with artificial atoms,
realizable in superconducting solid state systems. In those, artificial
two-level atoms with tunable eigenfrequencies have already been realized
\cite{Bouchiat1998,Nakamura1999}. Depending on the specific implementation, the
eigenfrequency can be changed by applying an external voltage or magnetic field.
But not only eigenfrequencies are tunable, there are also experiments where the
interaction strength between two artificial atoms can be varied, in absolute
value as well as in sign \cite{vanderPloeg2007,Harris2007}. These experiments
open up a broad range of in principle accessible parameters, which we want to
study in the following.
 
\section{Enhancement effect}
\label{Result}

\subsection{Starting parameters}
\label{StartingP}

Up to now we have left completely open on how to choose the eigenfrequencies and
coupling strengths to obtain an entangled steady state $\rho_{AB} = \tr_C\left[
\rho_{st} \right]$. Our simple line of guidance will be the case $J_C = 0$ where
the two systems $A$ and $B$ are only coupled to each other and their
environments, but not to the dissipative ancilla $C$. In this case the steady
state of system $A$ and $B$ alone can be obtained in a simple analytical form
\cite{Zhang2009,Sauer2014}, for which the Negativity, Eq. (\ref{negdef}), reads
\begin{equation}
 N\left( \rho_{AB} \right) = \textrm{max}\left[0,\frac{\sqrt{J^2 \gamma^2 +4 J^2
}-J^2}{4 J^2 + 4 + \gamma^2 } \right] . \label{2qneg}
\end{equation}
We will use this result to motivate our parameters. Remembering that we measure
all parameters in our system in units of $\omega$, we pick $\gamma = 10^{-3}$.
This seems reasonable, if we look at state of the art experiments
\cite{Dewes2012,Houck2009} where  eigenfrequencies of solid state qubits are
usually in the range of GHz, and at the same time decoherence parameters are
estimated to be in the lower MHz regime. 
This choice of parameters inserted into Eq. (\ref{2qneg}) tells us immediately,
that for a maximal entanglement of $N = \frac{1}{8}(\sqrt{5}-1) = 0.155$ we need
a coupling strength of $|J| = 0.62 $.
Hence the steady state entanglement in the $AB$ system should be observable in a
strong coupling regime \cite{Pashkin2003,Izmalkov2004,Majer2005}, where the
coupling strength $J$ is almost equal to unity.

This reasoning justifies the initial choice
\begin{equation}
 \gamma = 10^{-3} \ , |J| = 0.62
\end{equation}
of our parameters for which the steady state of $A$ and $B$ alone shows the
maximal bipartite entanglement of $N = 0.155$. Obviously, if we now switch on
the coupling $J_C$ with ancilla $C$, these parameters do not necessarily
describe the point of maximal entanglement between $A$ and $B$. Nonetheless this
is a valuable starting point from which we begin to study the influence of the
dissipative ancilla on the entanglement of our bipartite system. To asses our
improvements in entanglement, we recall the upper bound of $N=0.5$, realized by
a Bell state \cite{Nielsen2000}.

\subsection{Varying eigenfrequency and coupling strength of the dissipative
ancilla}
\label{efcs}

\begin{figure}[htb!]
 \includegraphics[width=0.99\columnwidth]{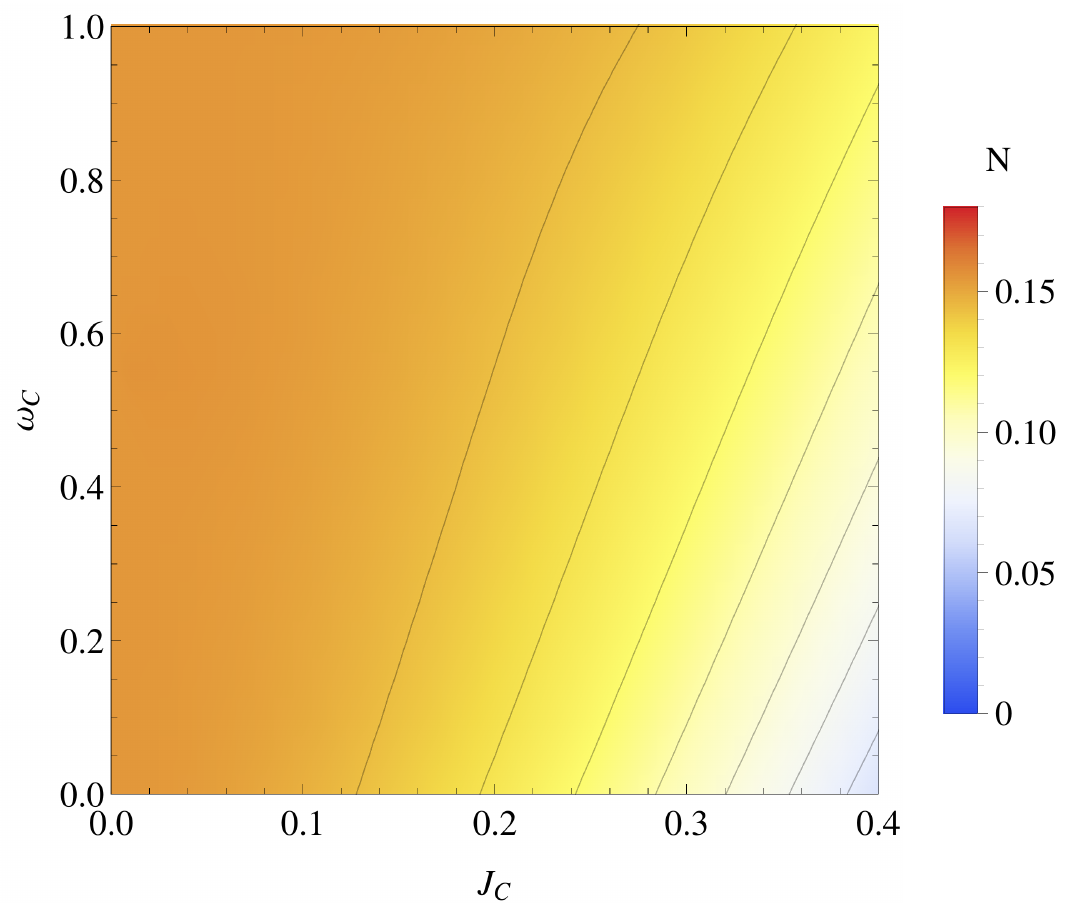}
 \caption{For $J = + 0.62$ and $\gamma_C = \gamma = 10^{-3}$ the steady state
entanglement between systems $A$ and $B$, as measured by the Negativity $N$,
decays with increased absolute value of coupling $J_C$ to the dissipative
ancilla $C$. The entanglement is maximal for $J_C = 0$ and arbitrary $\omega_C$,
this means the dissipative ancilla cannot enhance the bipartite entanglement
$N$.}
 \label{figJpos}
\end{figure}

Now we want to understand how the bipartite entanglement $N$, Eq.
(\ref{negdef}), between $A$ and $B$ varies as a function of the eigenfrequency
$\omega_C$ and the coupling strength $J_C$ of the ancilla $C$. At first, we keep
the decoherence parameter $\gamma_C = \gamma = 10^{-3}$ fixed. In Fig.
\ref{figJpos} we have chosen $J = + 0.62$ and observe a merely decaying
behavior of the entanglement $N$ with increasing coupling $J_C$ to the
dissipative ancilla. This is the somehow expected result: More decoherence in
the system reduces steady state entanglement.

\begin{figure}[htb!]
 \includegraphics[width=0.99\columnwidth]{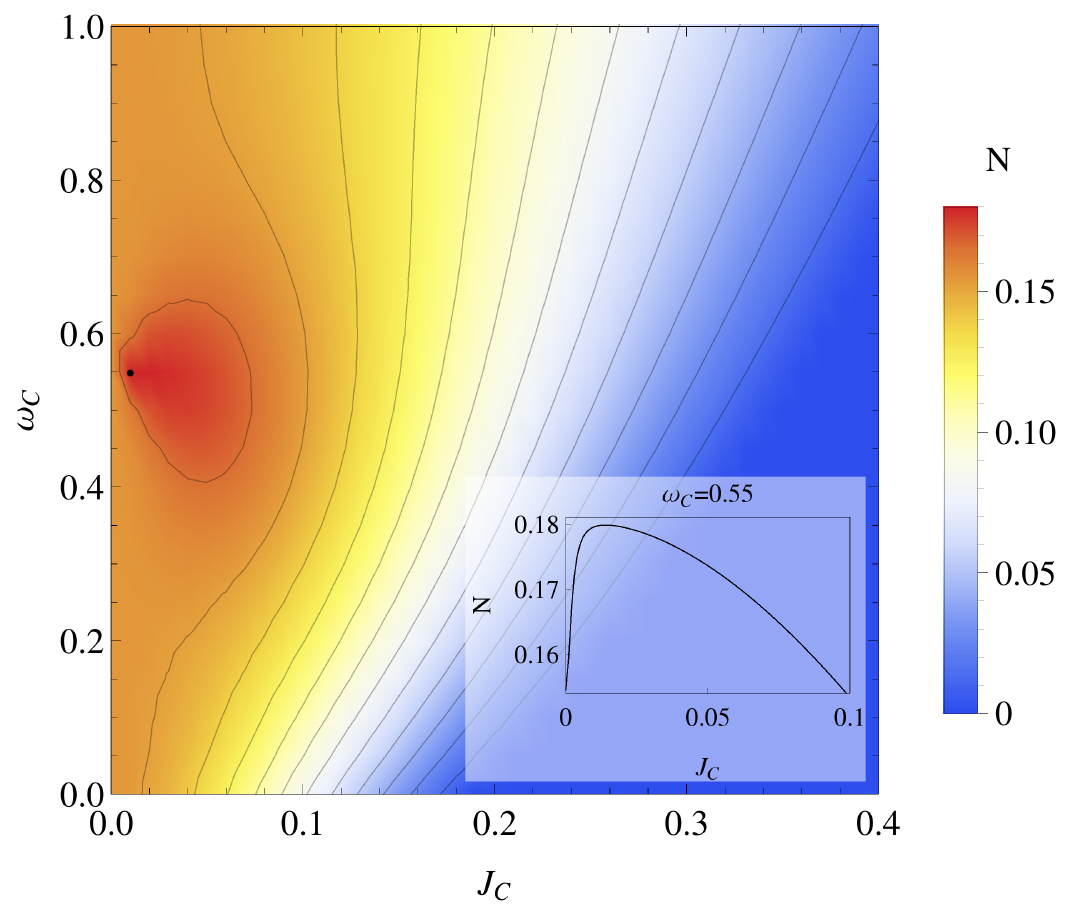}
 \caption{For $J = -0.62 $ and $\gamma_C = \gamma = 10^{-3}$ the entanglement
between systems $A$ and $B$, as measured by the Negativity $N$, does not simply
decrease when we increase the interaction strength $J_C$ with the dissipative
ancilla $C$. There is a distinctive maximum $N = 0.180$ (black dot) of
entanglement for an appropriately chosen eigenfrequency $\omega_C = 0.55$
and coupling strength $J_C =  0.01$, also visualized in the inset.}
 \label{figJneg}
\end{figure}

Next we study the case $J = - 0.62$. In Fig. \ref{figJneg} we show again the
bipartite entanglement $N$ between $A$ and $B$ as a function of the coupling
strength $J_C$ and eigenfrequency $\omega_C$ of the dissipative ancilla $C$,
keeping the decoherence parameter $\gamma_C = \gamma = 10^{-3}$ fixed. The
entanglement does not simply decrease when we increase the interaction strength
$J_C$, but shows a distinctive maximum of $N = 0.180$ (black dot in Fig.
\ref{figJneg}) for appropriately chosen eigenfrequency $\omega_C = 0.55$
and coupling strength $J_C = 0.01$. Interestingly, here the interaction
with the dissipative ancilla $C$ boosts the bipartite entanglement in the steady
state of system $AB$.

The only difference between Figs. \ref{figJpos} and \ref{figJneg}, regarding the
parameters used to calculate them, is the sign of the coupling strength $J$. 
It has been of no importance for the maximal entanglement in the uncoupled case
$J_C=0$, see Eq. (\ref{2qneg}). However, when we look at the full system $ABC$,
the sign of $J$ is crucial for the behavior of the steady state entanglement.
Unfortunately, in this case we have no similarly simple expression for the
Negativity as Eq. (\ref{2qneg}), telling us if the sign of a parameter is of
importance or not. A numerical study shows that the sign of the system-ancilla
coupling strength $J_C$ is of no importance, but to find local maxima in the
steady state entanglement $N$, it is necessary that the intra-system coupling
strength $J$ is negative.

This enhancement effect in bipartite entanglement occurs passively by just
adding the dissipative ancilla. We have no active driving elements, like an
external laser pumping a specific transition, in our system. Spontaneous decay
is present in all three subsystems $A$, $B$ and $C$. Yet, as in the case $J_C=0$
for an uncoupled bipartite system, the steady state $\rho_{AB}$ is still
entangled and this bipartite entanglement can even be enhanced. Through the
coupling of ancilla $C$ to our bipartite system $AB$, we obviously increase the
space of possible steady states, which can be reached dynamically. 
Moreover, we retain the full quantum character of this dynamics, as we trace out
the dissipative ancilla $C$ without further approximations.

\subsection{Decoherence parameter dependence}
\label{dpd}

\begin{figure}[htb!]
 \includegraphics[width=0.99\columnwidth]{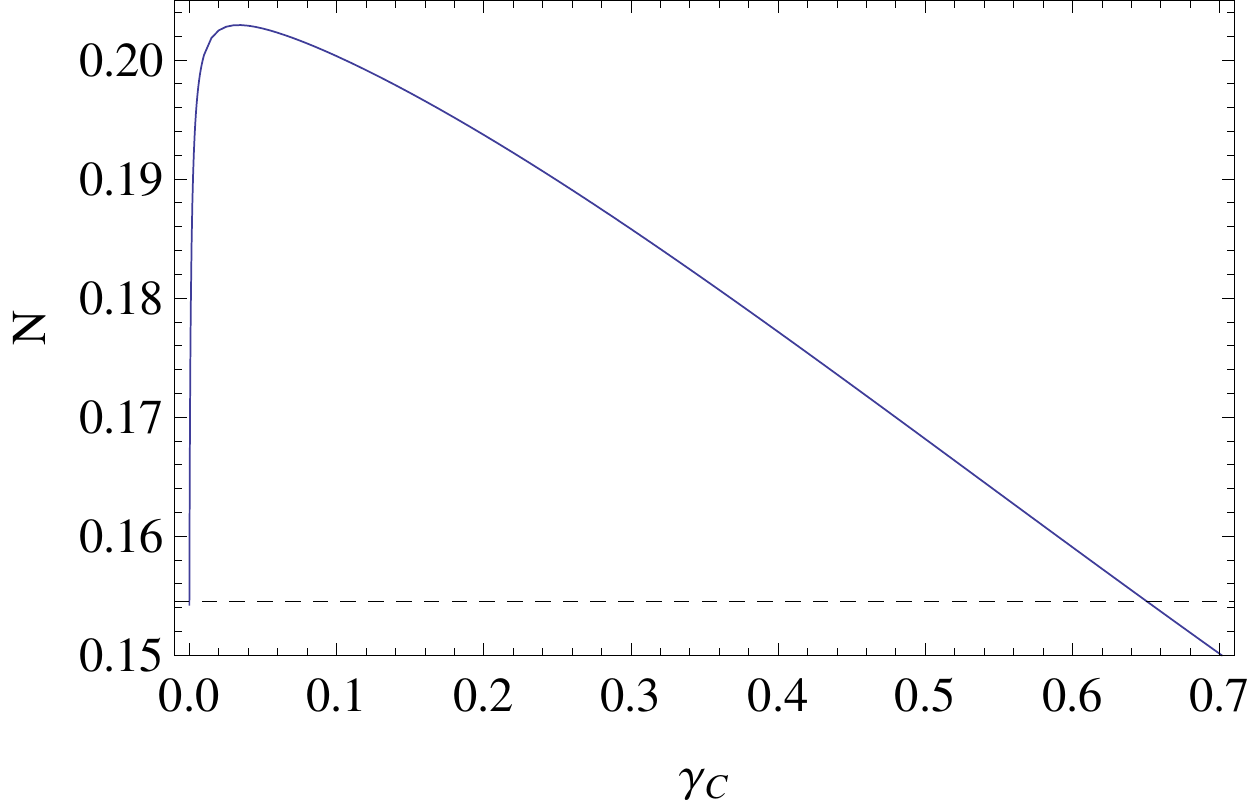}
 \caption{The value of bipartite entanglement $N$ in dependence of the
decoherence rate $\gamma_C$. The parameters $\omega_C = 0.55$, $J_C
=  0.01$, $J = - 0.62$ and $\gamma = 10^{-3}$ are kept constant. Only for
values of $\gamma_C > 0.64$ the enhancement effect ceases to exist and the
entanglement in the uncoupled system of $N=0.155$ (black dashed line) is
superior.}
 \label{gammaCN}
\end{figure}

In order to fully bring out the importance of the ancilla $C$ coupled to a bath
we turn the pike and fix the eigenfrequency $\omega_C = 0.55$ and the
coupling strength $J_C =  0.01$ while varying $\gamma_C$.
Now we investigate the dependence of the bipartite entanglement $N$ on the decay
rate $\gamma_C$ of our dissipative ancilla, see Fig. \ref{gammaCN}.
Starting form a small $\gamma_C$ the entanglement increases rapidly to an
extremal value of $N = 0.203$ at $\gamma_C = 0.04$ to then decrease again.
However, we find the enhanced entanglement over a wide range of dissipation
until we reach $\gamma_C > 0.64$, where the enhancement ceases to exist. In this
case the maximal entanglement may only be obtained by decoupling the dissipative
ancilla.
Nevertheless, in the large regime studied here, more decoherence, as measured by
$\gamma_C$, leads to more entanglement \footnote{Obviously, the position of the
maximal entanglement $N$ (black dot in Fig. \ref{figJneg}, for $\gamma_C =
10^{-3}$) is not fixed in the $(\omega_C, J_C)$ parameter space, if we vary the
decoherence parameter $\gamma_C$. If one determines the optimal values
$\omega_C$ and $J_C$ for every $\gamma_C$ dynamically, a curve can
be plotted that shows how the maximal entanglement $N$ depends on the
decoherence parameter $\gamma_C$. We omit this plot here, as it is qualitatively
equivalent to Fig. \ref{gammaCN}.} between $A$ and $B$.

In a realistic scenario, $\gamma_C$ is not a parameter to be engineered. In
contrast to the eigenfrequencies and coupling strengths, that may be tunable,
depending on the actual physical realization of the system
\cite{Bouchiat1998,Nakamura1999,vanderPloeg2007,Harris2007}, one cannot simply
change the rate of decoherence. Yet, one could think of methods to increase the
spontaneous decay of an artificial atom, for example by placing it in a more or
less resonant cavity, that is lossy itself. So, at least in principle, it should
be possible to influence $\gamma_C$ to some degree and thereby maximize the
entanglement enhancement effect observed here. On the other hand, even if the
decoherence parameter $\gamma_C$ may not be tuned, still Fig. \ref{gammaCN}
tells us that there is a broad range of possible decoherence parameters
$\gamma_C$ for which the enhancement effect should be observable.

Thus, arguing from a more fundamental point of view, our setup could also
be used to put the decoherence model included in the calculation of this quantum
effect to a test. In our model all three systems $A$, $B$ and $C$ are coupled to
individual heat baths. Their interaction in presence of environments leads to
the entanglement enhancement effect. If this is only an artifact of our model,
that does not appear in a real system, this model cannot be used to describe
intricate quantum effects of decoherence properly.

\section{Enhancement and the composition of the steady state}
\label{dsoe}

\subsection{Basic idea}

To understand the observed effect of entanglement enhancement further, we study
how the entangled steady state is dissipatively prepared
\cite{Poytas1996,Kraus2008,Verstraete2009,Schirmer2010}, starting from the
initial state
\begin{equation}
 \rho(t=0) = \ket{e}\bra{e}_A \otimes \ket{g}\bra{g}_B \otimes \ket{g}\bra{g}_C
\ , \label{inist}
\end{equation}
where subsystem $A$ is in the excited state $\ket{e}$, while $B$ and the
dissipative ancilla $C$ are in their ground states $\ket{g}$, respectively. 
Actually, the steady state solution $\rho_{st}$ does not depend on the initial
state and hence every choice will lead to the same result after more or less
rich dynamics. Here we have chosen just one generic example of an initial state
to exemplify how the steady state is reached in time.

In fact we will see that the dissipative dynamics selects a specific eigenstate
$\ket{E_n}$ of the undamped three-particle interaction, described by the
solution of
\begin{equation}
  \left( H_f + H_{int} \right) \ket{E_n} = E_n \ket{E_n} \ ,
\end{equation}
with Hamiltonians given by Eqs. (\ref{Hfree}) and (\ref{Hinteraction}). 
In order to trace this selection process in the course of time, we regard the
fidelities
\begin{equation}
 F_n(t) \equiv \bra{E_n} \rho(t) \ket{E_n}  \label{fidelities}
\end{equation}
of the time evolved state $\rho(t)$, Eq. (\ref{louivillian}), and the
eigenstates $\ket{E_n}$.

\begin{figure}[htb!]
  \subfigure[\quad $J = +0.62, \gamma_C = 10^{-3}$ \label{es1a}]{
\includegraphics[width=0.99\columnwidth]{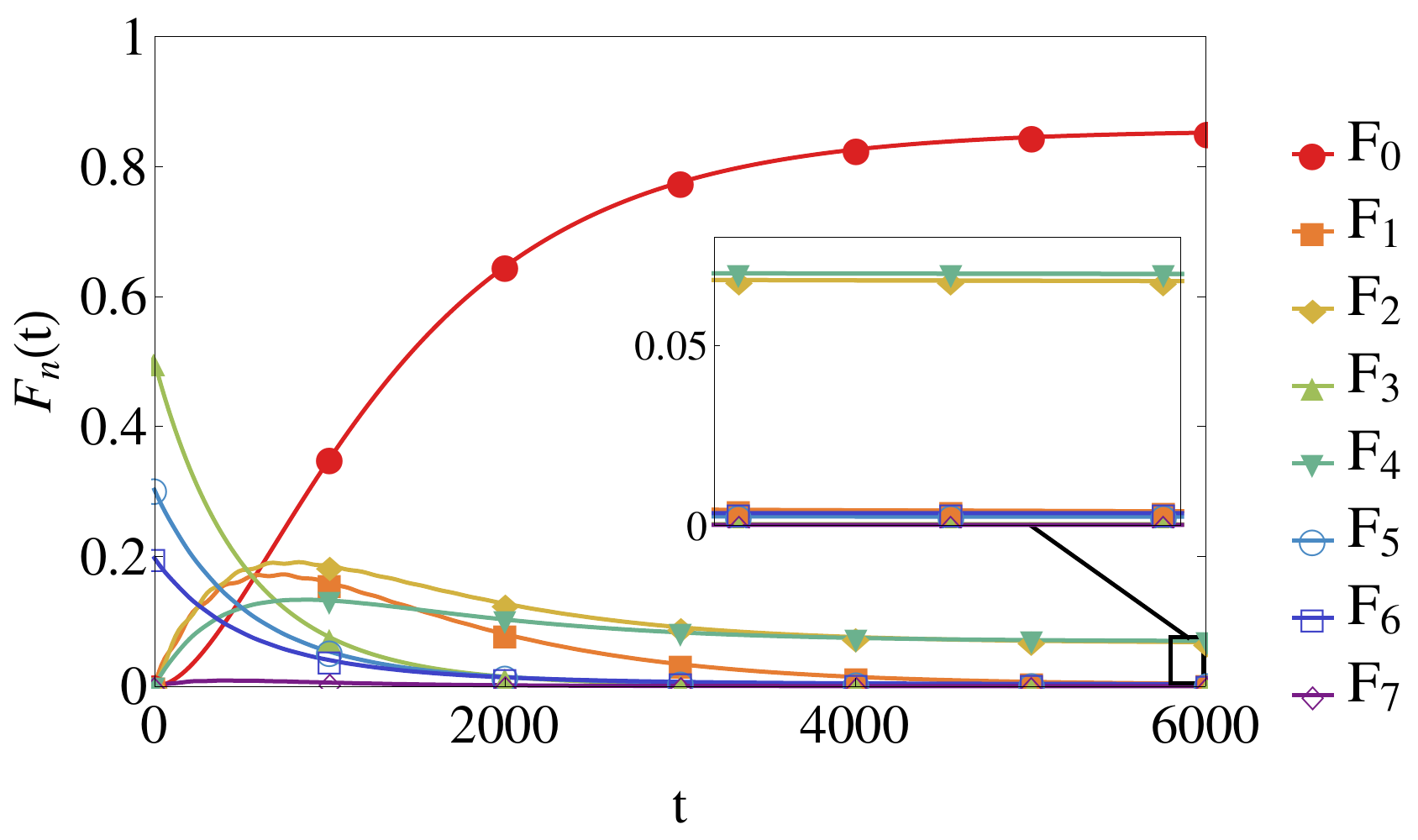}}
  \subfigure[\quad $J = -0.62, \gamma_C = 0.04$ \label{es1b}]{
\includegraphics[width=0.99\columnwidth]{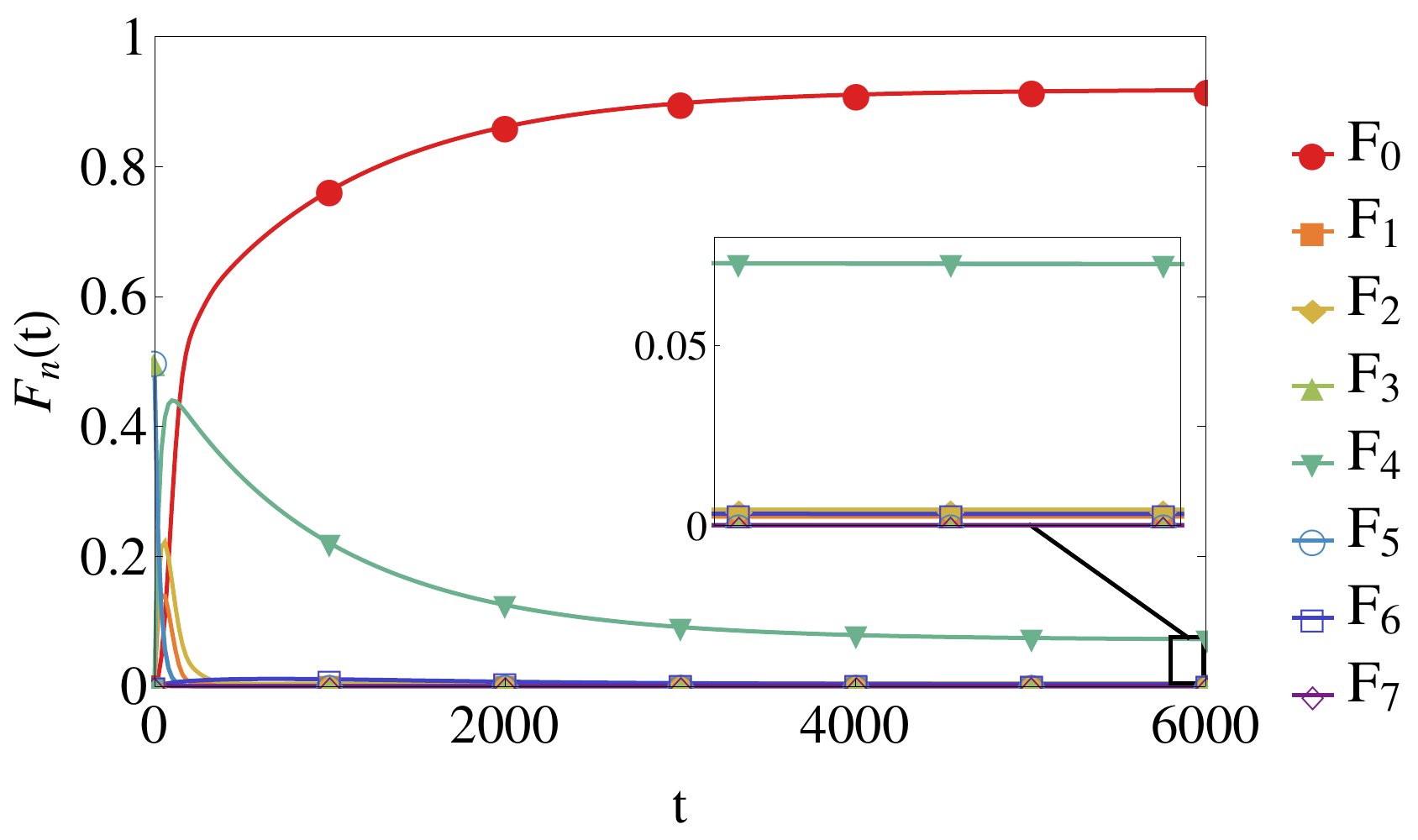}}
  \caption{We show the dynamics of the fidelities $F_n(t)$, starting from the
initial state in Eq. (\ref{inist}). In particular, we compare the parameter sets
from Eqs. (\ref{esparaI}) and (\ref{esparaII}) in Figs. \ref{es1a} and
\ref{es1b}, respectively. We observe in both, that after a transient evolution
an equilibrium is reached, where the mixture of the steady state is dominated by
the individual ground state $\ket{E_0}$. The insets illustrate, that by changing
the sign of $J$ and choosing the optimized decoherence parameter $\gamma_C$ the
mixture is mainly altered by suppressing state $\ket{E_2}$ at the expense of
state $\ket{E_0}$. This explains the change in entanglement with respect to
subsystem $AB$.}
  \label{es1}
\end{figure}

The natural expectation would be that a dissipative time evolution finally
drives the three-particle system into its ground state $\ket{E_0}$.
This is what we basically observe in Fig. \ref{es1a} for the parameter set
\begin{equation}
J = 0.62, \ \gamma_C=\gamma=10^{-3}, \ \omega_C = 0.55, \  J_C = 0.01 ,
\label{esparaI}
\end{equation}
representing the parameter regime also depicted in Fig. \ref{figJpos} of the
previous section.
Starting from the initial state, Eq. (\ref{inist}), we first observe a transient
dynamics, which afterwards merges into the steady state solution. The special
fact to be observed in Fig. \ref{es1a} is the mixture in the steady state, as
depicted in the inset. We see that besides the expected major contribution of
the corresponding ground state $\ket{E_0}$, we still have a small
fraction of higher lying eigenstates. Hence we can approximate the steady state
density operator by
\begin{equation}
 \rho_{st} \approx \sum\limits_{n \in \{ 0 ,2 ,4 \} } F_n \ \ket{E_n} \bra{E_n}
\ ,
\end{equation}
which carries the Negativity $N(\rho_{AB})=0.157$ for the $AB$ system in state
$\rho_{AB}= \tr_C[\rho_{st}]$ and hence nicely approximates the typical
Negativities shown in Fig. \ref{figJpos}.

If we now choose the parameter set
\begin{equation}
  J = - 0.62, \  \gamma_C = 0.04, \ \omega_C= 0.55, \ J_C =  0.01, \
\gamma=10^{-3},  \label{esparaII}
\end{equation}
for which we observed the maximal enhancement effect in the previous section, we
recognize a very similar transient dynamics in Fig. \ref{es1b}, leading to a
strong contribution of the ground state $\ket{E_0}$ \footnote{Notice that by
changing parameters that appear in the Hamiltonian, the according eigenstates
change as well. As it is clear from context which set of parameters we use and
to simplify notation, we abstain from introducing additional indexes.}. However,
the selection of the higher energy eigenstates changes: $\ket{E_2}$ is
suppressed at the expense of $\ket{E_0}$, but the state $\ket{E_4}$ survives,
so that we arrive at the approximate density operator
\begin{equation}
 \rho_{st} \approx \sum\limits_{n \in \{ 0 ,4 \} } F_n \ \ket{E_n} \bra{E_n} \ ,
\end{equation}
which indeed leads to the enhanced Negativity $N(\rho_{AB})=0.206$, seen in Fig.
\ref{gammaCN}.
Obviously, the change in composition of energy eigenstates in the steady state,
explains why entanglement is enhanced by dissipatively coupling an ancilla to
systems $A$ and $B$ with a suitable choice of parameters.
However, the enhancement is still rather small since the ground state
$\ket{E_0}$ dominates this composition. The question then is whether we can even
select an eigenstate, like the surviving state $\ket{E_4}$, with a much higher
degree of $AB$ entanglement by choosing the right set of parameters. Thus we can
try to look at the problem from a fully engineering perspective in the next
section. 

\subsection{Optimum}

In principle, all eigenfrequencies and coupling
strengths, and with limitations also the decoherence parameters, are adjustable.
In other words, the steady state of the full Liouvillian, Eq.
(\ref{louivillian}), is a function of the parameters $J, \gamma, \omega_C, J_C$
and $\gamma_C$. As explained in section \ref{StartingP}, with regard to
experiment, we fix the reasonable choice $\gamma = 10^{-3}$. Next we constrain
the remaining parameters by
\begin{eqnarray}
  -1 \leq J \leq 0, \quad & 0 \leq J_C \leq 1, \nonumber \\
  -1 \leq \omega_C \leq 1, \quad &  0 < \gamma_C \leq 1 . 
\end{eqnarray}
These constraints are justified, as we consider similar two-level systems with
comparable eigenfrequencies and decoherence parameters. Also, to stay in a
physically reasonable parameter regime, the interaction strengths are bounded.

\begin{figure}[htb!]
 \includegraphics[width=0.99\columnwidth]{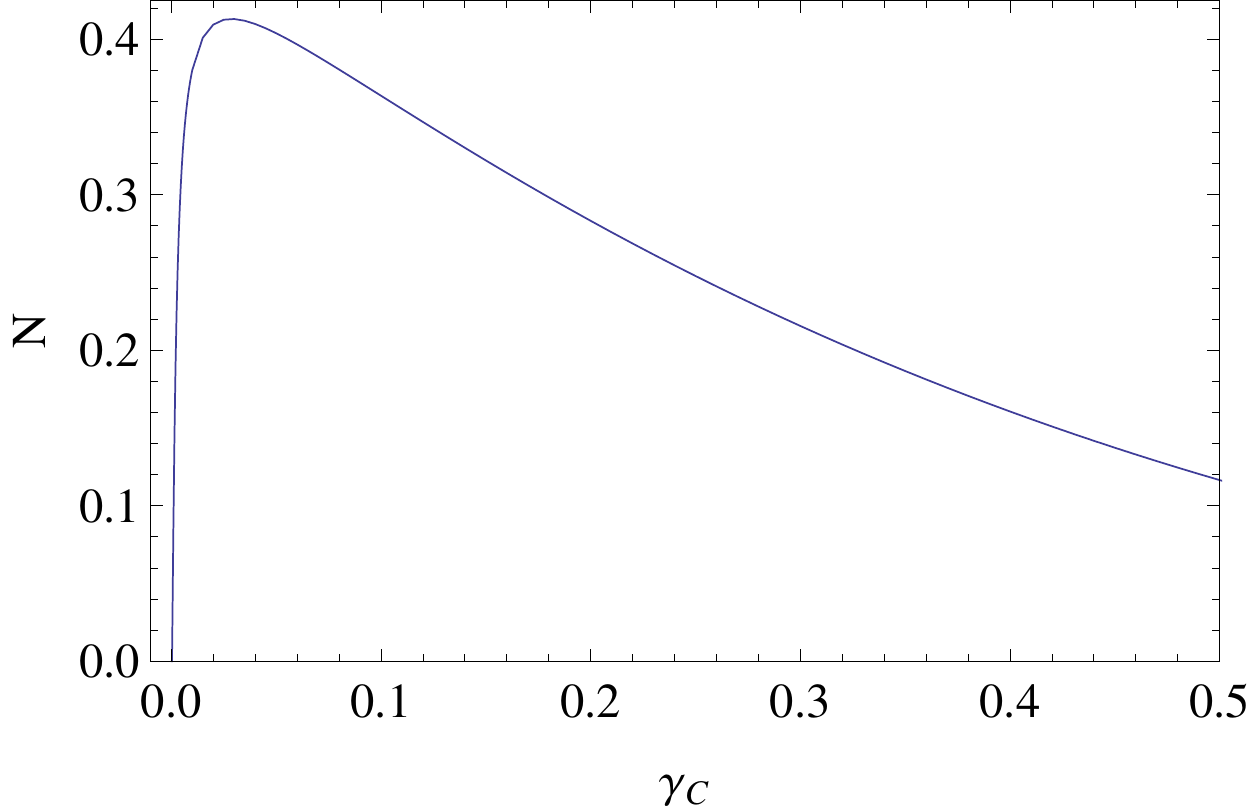}
 \caption{The value of bipartite entanglement $N$ in dependence of the
decoherence rate $\gamma_C$. The parameters $\omega_C = -0.74$, $J_C = 0.01 $,
$J = -0.31$ and $\gamma = 10^{-3}$ are kept constant. As seen before in Fig.
\ref{gammaCN}, the bath of ancilla $C$ is crucial to obtain an enhancement
effect.}
\label{04gammaCN}
\end{figure}

Numerically solving this constrained optimization problem, i.e. finding the
steady state solution of Eq. (\ref{louivillian}) with maximal bipartite
entanglement $N$ between systems $A$ and $B$, yields a maximum of 
\begin{equation}
 N_{max} = 0.413
\end{equation}
for the optimal parameter set
\begin{equation}
 J = -0.31, \gamma_C = 0.03, \omega_C= -0.74, J_C = 0.01, \gamma = 10^{-3}. 
\label{optpara}
\end{equation}
The maximal entanglement $N_{max} = 0.413$ is quite an improvement over the
maximal entanglement of $N=0.155$ in the uncoupled case.
At the same time it is again crucial that the ancilla is dissipative, see Fig.
\ref{04gammaCN}. 
The maximum at $\gamma_C = 0.03$ is surrounded by broad sides, where the
entanglement is still enhanced. For very small and large values of the
decoherence parameter $\gamma_C$, the entanglement drops below $N=0.155$. Hence,
we need just the right coupling to a bath for ancilla $C$ to achieve an
enhancement effect.

\begin{figure}[htb!]
 \includegraphics[width=0.99\columnwidth]{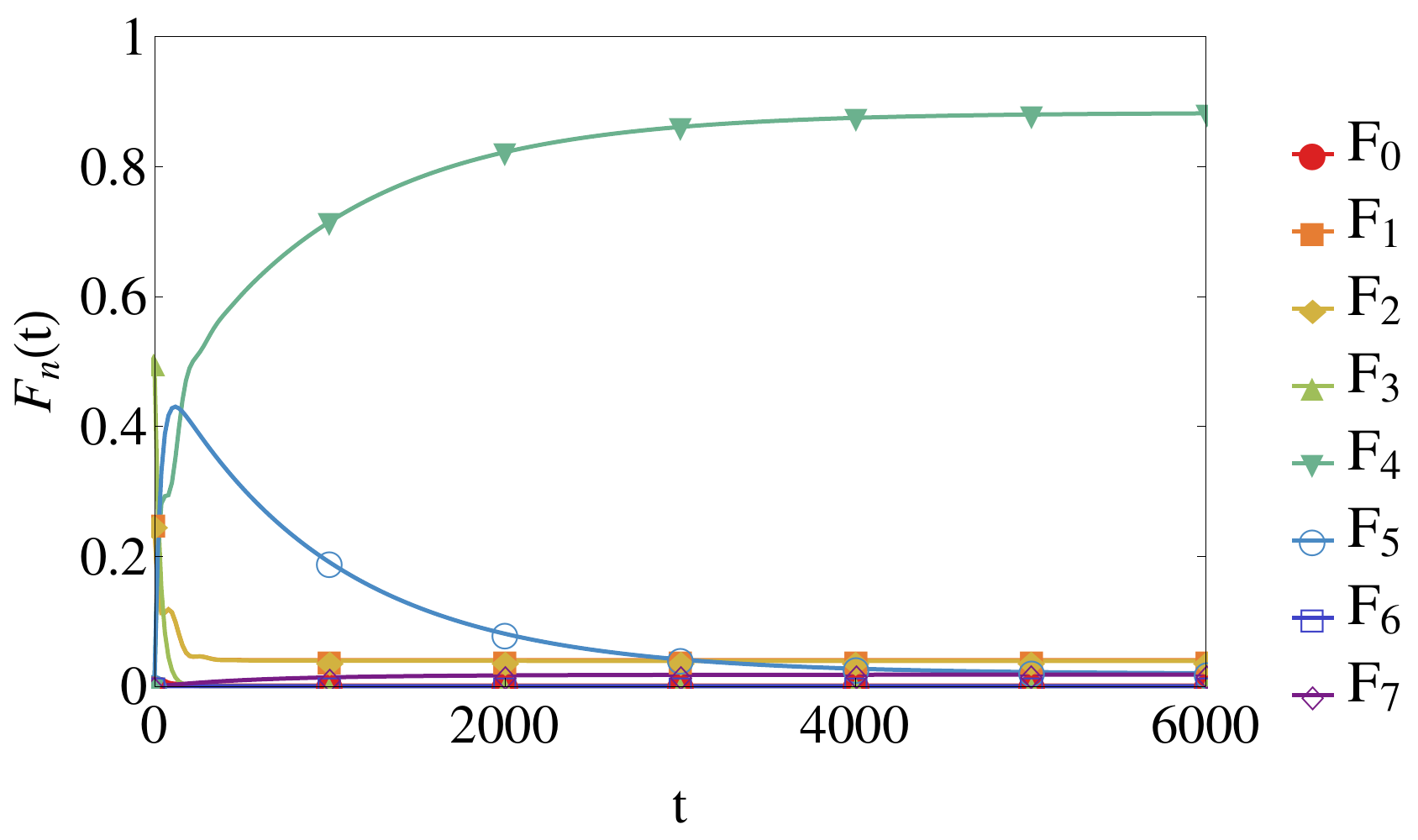}
 \caption{The dynamics of the fidelities $F_n(t)$, starting from the initial
state in Eq. (\ref{inist}) with the parameters from Eq. (\ref{optpara}). After a
transient evolution an equilibrium is reached where the mixture of the steady
state is now dominated by the excited state $\ket{E_4}$. This is in stark
contrast to the situation observed in Fig. \ref{es1} and explains the
significantly different entanglement properties.}
\label{es2}
\end{figure}

As already mentioned in the previous section, this improvement in entanglement
can be explained in more detail, if we once again look at the dynamical
selection of eigenstates for $t \rightarrow \infty$.
In Fig. \ref{es2} we show how the fidelities $F_n(t)$, Eq. (\ref{fidelities}),
evolve in time, again starting from the initial state in Eq. (\ref{inist}). 
Different than before this time not the ground state $\ket{E_0}$ dominates the
mixture, but the excited state $\ket{E_4}$. This is clearly visible from the
dynamics, where one can see $F_4$ rising to its steady state value. As the
eigenstate $\ket{E_4}$ possesses bipartite entanglement of  $N = 0.499$ in
subsystem $AB$, it is possible to understand why the entanglement is boosted to
such a level at this point in parameter space. Here a totally different
eigenstate of the Hamiltonian is selected and responsible for the enhancement in
bipartite entanglement. 

Compared to the maximal possible bipartite entanglement of $N=0.5$, which is
realized by a Bell state, the enhancement that results from coupling systems $A$
and $B$ with an specifically engineered ancilla $C$ is remarkable.

\section{Conclusion}
\label{Conclusion}

In this manuscript we have studied how to enhance the steady state entanglement
between two two-level systems, by coupling them to a dissipative ancilla. This
ancilla has also been chosen to be a two-level system, subject to the same
spontaneous decay noise like the other systems. We have found an enhancement
effect in entanglement, depending on the coupling strength, eigenfrequency and
decoherence parameter of the ancilla. As we have no active driving elements
present, the enhancement occurs passively. Also, the interaction with the
dissipative ancilla has been treated without any approximations, i.e. the result
contains the full quantum dynamics of the interaction between system and
ancilla. We have shown that the enhancement effect is intimately connected to
the composition of the steady state, that we expressed in terms of eigenstates
of the undamped three-particle system. Coupling a dissipative ancilla has
allowed us to alter this composition significantly, enabling the enhancement of
entanglement.

The optimal parameters, determined at last, show that a remarkably large
enhancement is possible. The associated mixed steady state represents the
optimized result of a dissipative state preparation, with respect to the
restriction of fixed coupling and Lindblad operators. Since we have started
from an experimentally motivated model, a realization of a related setup might
be within reach, establishing the possibility to exploit the observed
enhancement effect to engineer entangled states.

Moreover, an experimental approach will shed additional light on our theoretical
concepts for the description of decoherence. The more or less phenomenological
modeling of spontaneous decay as primary source of decoherence is crucial for
the appearance of the enhancement effect. In an experiment, this model would be
put to a sensitive test. Beyond that, extensions of the decoherence model, for
example regarding finite temperature heat baths or adding additional dephasing
noise, are possibilities of further studies. 

In addition, due to the model's simplicity and the symmetries involved, at least
in some regimes an analytical treatment might be within reach. This could for
example be used to study, if there is an even more fundamental mechanism behind
the selection of eigenstates explaining the enhancement effect.

\end{document}